\documentclass[preprint,12pt]{elsarticle}

\pdfoutput=1

\usepackage{epsfig}
\usepackage{graphics}
\usepackage{graphicx}
\usepackage[centertags]{amsmath}
\usepackage{amsfonts}
\usepackage{amsthm}
\usepackage{amssymb}
\usepackage{url}
\usepackage{subfigure}

\newtheorem{proposition}{Proposition}

\newtheorem{assumption}{Assumption}

\begin{document}

%\begin{frontmatter}

\title{On computing in fine-grained \\ compartmentalised Belousov-Zhabotinsky medium}

\author{Andrew Adamatzky, Julian Holley, Larry Bull, Ben De Lacy Costello}

\address{University of the West of England, Bristol, United Kingdom}

\date{\today}

\begin{abstract}

\vspace{0.5cm}

\noindent
We introduce results of computer experiments on information processing
in a hexagonal array of vesicles filled with Belousov-Zhabotinsky (BZ)
solution in a sub-excitable mode. We represent values of Boolean
variables by excitation wave-fragments and implement basic logical
gates by colliding the wave-fragments. We show that a vesicle filled
with BZ mixture can implement a range of basic logical functions. We
cascade BZ-vesicle logical gates into arithmetic circuits implementing
addition of two one-bit binary numbers.  We envisage that our
theoretical results will be applied in chemical laboratory designs of
massive-parallel computers based on fine-grained compartmentalisation
of excitable chemical systems.

\vspace{0.5cm}

\end{abstract}

%\pacs{82.40.-g; 82.40.Ck; 89.75.Kd; 89.75.Fb; 89.20.Ff}

\maketitle

%\end{frontmatter}

\section{Introduction}

A reaction-diffusion computer is a spatially extended chemical system
which processes information using interacting growing patterns of
excitable and diffusive
waves~\cite{adamatzky_delacycostello_asai_2005}. In reaction-diffusion
processors, both the data and the results of the computation are
encoded as concentration profiles of the reagents. The computation is
performed via the spreading and interaction of wave fronts. All
reaction-diffusion computers can be roughly split into two types:
geometrically-constrained and
free-space~\cite{adamatzky_privman_issue}. In
geometrically-constrained computers excitation or chemical waves
propagate in the `hardware' of substrate defined channels and interact
at the junctions between the channels. A great deal of experimental
reaction-diffusion computing circuits were implemented in a
geometrically-constrained media: logical gates for Boolean and
multiple-valued logic~\cite{sielewiesiuk_2001, motoike_2003,
  gorecki_2009, yoshikawa_2009}, many-input logical
gates~\cite{gorecki_2006,gorecki_2006a}, counters~\cite{gorecki_2003},
coincidence detector~\cite{gorecka_2003}, detectors of direction and
distance to a source of periodic
oscillations~\cite{gorecki_2005,yoshikawa_2009a} and inductive
memory~\cite{motoike_2003}.

Free-space reaction-diffusion computers\footnote{The term `free-space computing' is coined by
  Jonathan Mills} do not have any underlying architecture.  A
computing medium is uniform and homogeneous, the only externally
evoked disturbances of the medium's characteristics implement
computation. In majority of cases computation in free-space computers
is executed using principles of a collision-based
computation~\cite{adamatzky_cbc}. A collision-based computer employs
mobile compact finite patterns and mobile self-localized excitations
to represent quanta of information in active non-linear mediums.
Information values, e.g. truth values of logical variables, are given
by either absence or presence of the localizations or other parameters
of the localizations.  The localizations travel in space and when
collisions occur the result can be interpreted as computation. There
are no predetermined stationary signal channels (wires), a trajectory
of the travelling pattern is a momentarily wire. Almost any part of
the medium space can be used as a wire. Localizations can collide
anywhere within a space sample, there are no fixed positions at which
specific operations occur, nor location specified gates with fixed
operations. The localizations undergo transformations, form bound
states, annihilate or fuse when they interact with other mobile
patterns. Information values of localizations are transformed as a
result of collision~\cite{adamatzky_cbc}.

So far no advanced arithmetical circuits have been implemented in
free-space reaction-diffusion chemical processors.  Some preparatory
steps are done however. They include simulation and experimental
laboratory realisation of basic logical
gates~\cite{adamatzky_2004_collision,andy_ben_BZ_collision,RITABEN}
and generators of mobile localizations, excitation wave-fragments
(they can play a role of constant {\sc True}) in light-sensitive
Belousov-Zhabotinsky (BZ) medium~\cite{ben_gun} and adaptive design
of simple logical gates using machine learning techniques~\cite{toth_2009}.    
The main reason for such slow progress is instability of wave-fragments that either
collapse or expand without external control.  One-bit half-adder in
geometrically-constrained light-sensitive Belousov-Zhabotinsky medium
was simulated in~\cite{adamatzky_physarumgates}, however the
implementation required dynamical update of illumination level to
prevent wave-fragments from collapsing or expanding. In this present
paper we try to overcome the problem of excitation wave-fragments
instability via combining geometrically-constraining and
collision-based approaches.
 
The paper is structured as follows. Two-variable Oregonator model of
BZ medium is introduced in Sect.~\ref{methods}.
Section~\ref{compartmentalisation} outlines principles of
compartmentalisation of BZ medium in planar discs and packing of the
discs into a two-dimensional hexagonal lattice. Typology of
interactions between wave-fragments is presented in
Sect.~\ref{interaction}. In Sect.~\ref{CAmodel} we introduce a
cellular-automaton model of interacting wave-fragments.  We show what
types of logical gates can be realised via collision of wave-fragments
in Sect.~\ref{gatessection}. These gates are employed to construct a
one-bit full adder in Sect.~\ref{adder}. Future developments of the
approach are outlined in Sect.~\ref{discussion}.

\section{Oregonator model}
\label{methods}

We use two-variable Oregonator equation~\cite{field_noyes_1974} adapted to a light-sensitive 
Belousov-Zhabotinsky (BZ) reaction with applied
illumination~\cite{beato_engel}.

\begin{eqnarray}
  \frac{\partial u}{\partial t} & = & \frac{1}{\epsilon} (u - u^2 - (f v + \phi)\frac{u-q}{u+q}) + D_u \nabla^2 u \nonumber \\
  \frac{\partial v}{\partial t} & = & u - v 
\label{equ:oregonator}
\end{eqnarray}

In framework of BZ reaction the variables $u$ and $v$ represent local
concentrations of activator, or excitatory component, and inhibitor,
or refractory component. Parameter $\epsilon$ sets up a ratio of time
scale of variables $u$ and $v$, $q$ is a scaling parameter depending
on rates of activation/propagation and inhibition, $f$ is a
stoichiometric coefficient. Constant $\phi$ is a rate of inhibitor
production. In light-sensitive BZ $\phi$ represents rate of inhibitor
production proportional to intensity of illumination
(\ref{equ:oregonator}).

To integrate the system we use Euler method with five-node Laplace
operator, time step $\Delta t=0.001$ and grid point spacing $\Delta x
= 0.25$, $\epsilon=0.0243$, $f=1.4$, $q=0.002$. In some cases it was
enough to speed up simulation by increasing $\Delta t$ to $0.005$
without loss of phenomenological accuracy.  The parameter $\phi$
characterizes excitability of the simulated medium: the medium is
excitable and exhibits `classical' target waves when $\phi=0.05$ and
the medium is sub-excitable with propagating localizations, or
wave-fragments, when $\phi=0.0766$. Time lapse snapshots provided in
the paper were recorded at every 150 time steps, we display sites with
$u >0.04$.

\section{Compartmentalisation}
\label{compartmentalisation}

\begin{figure}[!htb]
\centering
\subfigure[]{\includegraphics[angle=90,width=0.7\textwidth]{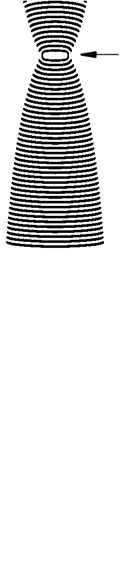}}
\subfigure[]{\includegraphics[angle=90,width=0.7\textwidth]{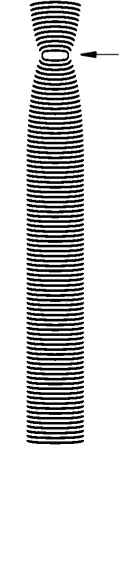}}
\subfigure[]{\includegraphics[angle=90,width=0.7\textwidth]{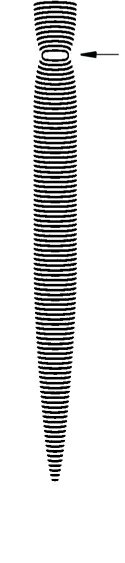}}
\caption{Time lapsed snapshots of wave-fragments propagating in
  simulated BZ medium in (a)~near lower threshold of excitability, $\phi=0.079$,
  (b)~sub-excitable mode, $\phi=0.07905$, (c)~non-excitable mode,
  $\phi=0.0791$. The media were perturbed by rectangular north-south
  elongated domains of excitation, $3 \times 40$ sites in state
  $u=1.0$.  Sites of initial segment-wise perturbation are shown by
  arrows. Grid size is 1125$\times$250 nodes. 
  Time lapsed snapshots are recorded every $150^{th}$ step of
  numerical integration.}
\label{fragments}
\end{figure}

Given initial asymmetric excitation domain a wave-fragment is
formed. The fragment's velocity vector is a normal to longest side of
the perturbation domain. In Fig.~\ref{fragments} a medium is excited
by rectangular domains of perturbed sites. The perturbation domains
are elongated along north-south axis therefore wave-fragments
generated propagate west and east. Depending on medium's excitability
wave-fragments may expand (Fig.~\ref{fragments}a), keep their shape
for a long time (Fig.~\ref{fragments}b) or collapse
(Fig.~\ref{fragments}c).

In ideal situation, assuming that wave-fragments keep their shapes
indefinitely, we can implement a collision-based computing circuit of
any depth subject to space availability.  However in reality,
particularly in conditions of chemical laboratory experiments,
wave-fragments are very unstable. It is almost impossible to keep a
medium at the precise level of sub-excitability
(Fig.~\ref{fragments}b), and almost any wave-fragment will expand
(Fig.~\ref{fragments}a) or collapse (Fig.~\ref{fragments}c). The only
solution, for controlling shape of wave-fragments, suggested so far
was to change excitability of a medium periodically, as achieved in
\cite{sakurai_2002} in experiments with light-sensitive BZ
medium. When a wave-fragment expands experimenters increase
illumination level thus decreasing the medium's excitability. When the
wave-fragment starts to collapse the illumination is decreased, the
medium's excitability increases and the wave-fragment expands.  This
approach does not look feasible for collision-based
computing~\cite{adamatzky_cbc} with excitation wave-fragments, because
it requires precise synchronisation of cycles of decreasing and
increasing excitability.

We can overcome the problem of wave-fragment instability by
the subdividing computing substrate into interconnected compartments ---
BZ-vesicles --- and allowing waves to collide one with another only
inside the compartments. Each BZ-vesicle has a membrane impassable for
excitation. A pore, or a channel, between two vesicles is formed when
two vesicles come into direct contact.  The pore is small such that
when a wave passes through the pore there is not enough time for a
wave to expand or collapse before interacting with waves entering
through other pores, or sites of contact.

\begin{figure}[!tbp]
\centering
\subfigure[]{\includegraphics[width=0.3\textwidth]{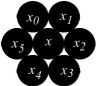}}
\subfigure[]{\includegraphics[width=0.6\textwidth]{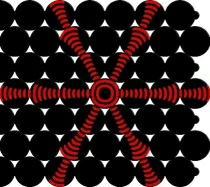}}
\caption{Basic architecture element of BZ compartmentalized medium:
  (a)~structure of disc-compartment neighbourhood, (b)~excitation
  propagation on a regular arrangement of vesicles.}
\label{ballsexample}
\end{figure}

A spherical compartment --- BZ-vesicle --- is the best natural
choice. It also conforms to experimental results on encapsulating
excitable medium in a lipid membrane~\cite{gorecki_private,neuneu} and
allows for effortless arrangement of the vesicles into a regular
lattice.

In this present paper we deal with a two-dimensional medium where
vesicles are tightly packed into a hexagonal lattice
(Fig.~\ref{ballsexample}a). Each vesicle has radius 27 grid units, a
pore has radius 5 units. The size of the vesicle and pore are critical
in modulating the beam of the excitation waves. Changes in vesicle size 
will require further adjustment of parameter
$\phi$ (Eq.~\ref{equ:oregonator}) in order to maintain the ideal wave
spread within the vesicle.

Each internal vesicle $x$ has six closest
neighbours $x_0 \cdots x_5$. Medium inside BZ-vesicles is
sub-excitable while the medium outside the vesicles are
non-excitable. Excitation wave-fragments propagate inside vesicles
only. The waves pass from one vesicle through pores, sites of
contacts.

\vspace{0.5cm}

\begin{proposition}
  BZ-vesicles packed in a hexagonal lattice provide a medium for
  directional transmission of information in a spatially extended
  non-linear excitable medium, where each vesicle acts as a signal
  amplifier and a pore, connecting two vesicles, acts as a focusing
  lens.
\end{proposition}

We excite the medium inside one vesicle, say $x$, with a symmetric
disc-shaped domain of perturbation centered in one of the vesicles. A
circular excitation wave is formed. The circular wave reaches boundary
of vesicle $x$, and enters pores between $x$ and its neighbours $x_0
\cdots x_5$ (Fig.~\ref{ballsexample}b). When excitation leaves a pore
it has a form of a propagating wave-fragment.  Due to the
near-excitability of the medium (Fig.~\ref{fragments}a) wave-fragments
entering vesicles $x_0 \cdots x_5$ start to expand. However they do
not have enough space to form a proper circular wave. Therefore they
leave second-order neighbours of $x$ via one exit pore per vesicle
(Fig.~\ref{ballsexample}b).

\section{Interaction between wave-fragments}
\label{interaction}

\begin{assumption}
  All constructs presented in the paper assume system of BZ-vesicles
  is fully synchronized. Waves enter any single vesicle
  simultaneously.
\end{assumption}

This is a very strong and somewhat unrealistic assumption. However we
use it as a starting point to build first reliable models of
BZ-vesicle computers.

\begin{figure}[!tbp]
\centering
\subfigure[]{\includegraphics[width=0.24\textwidth]{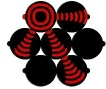}}\\
\subfigure[]{\includegraphics[width=0.24\textwidth]{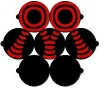}}
\subfigure[]{\includegraphics[width=0.24\textwidth]{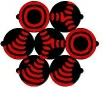}}
\subfigure[]{\includegraphics[width=0.24\textwidth]{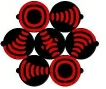}}\\
\subfigure[]{\includegraphics[width=0.24\textwidth]{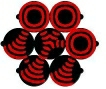}}
\subfigure[]{\includegraphics[width=0.24\textwidth]{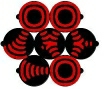}}
\subfigure[]{\includegraphics[width=0.24\textwidth]{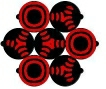}}
\subfigure[]{\includegraphics[width=0.24\textwidth]{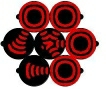}}
\subfigure[]{\includegraphics[width=0.24\textwidth]{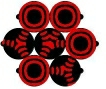}}
\subfigure[]{\includegraphics[width=0.24\textwidth]{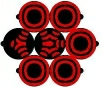}}
\subfigure[]{\includegraphics[width=0.24\textwidth]{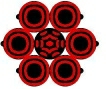}}
\caption{Time lapse trajectories of wave interactions in BZ-vesicles. Representative 
scenarios are given for (a)~one input wave, (b)--(d)~two input waves, (e)--(g)~three input waves,
(h)--(i)~four input waves, (j)~five input waves, and (k)~six input waves.}
\label{typesofwavecollisions}
\end{figure} 

\vspace{0.5cm}

\begin{proposition}
  Let several wave-fragments enter a vesicle. If at least two
  wave-fragments have opposite velocity vectors all wave-fragments
  annihilate. Otherwise the wave-fragments merge and velocity vector
  of newly formed wave-fragment is a sum of velocity vectors of
  incoming wave-fragments.
\end{proposition}

Constructive proof is illustrated by representative scenarios of wave
interactions inside a single vesicle (vesicle $x$ in
Fig.~\ref{ballsexample}a) are shown in
Fig.~\ref{typesofwavecollisions}.  If just one neighbour of
vesicle $x$ is excited the vesicle $x$ acts as a conductor and signal
amplifier: the wave simply passes through the vesicle $x$ slightly
increasing in size and exits through the pore opposite to the wave's
entry pore. Thus, in Fig.~\ref{typesofwavecollisions}a north-west
neighbours (vesicle $x_0$ in Fig.~\ref{ballsexample}a) is
activated. Excitation enters vesicle $x$, wave-fragment travelling
south-east is formed (Fig.~\ref{typesofwavecollisions}a). The
wave-fragment is transmitted to south-east neighbour (vesicle $x_3$ in
Fig.~\ref{ballsexample}a) of vesicle $x$.

There are three possible scenarios when two neighbours of vesicle $x$
are excited. In a situation when excited neighbours of $x$ are also
each others closest neighbours (e.g. vesicles $x_0$ and $x_1$ in
Fig.~\ref{ballsexample}a) the wave-fragments generated by them merge
inside vesicle $x$ (Fig.~\ref{typesofwavecollisions}b). The velocity
vector of a newly formed wave-fragment is a sum of vectors of two
original wave-fragments. Vectors of original wave-fronts orientate
towards exit pores opposite to the excitation entry pores. The vector
of a newly formed wave-fragment aims between the pores, hence no
excitation leaves the vesicle $x$. For example, in
(Fig.~\ref{typesofwavecollisions}b) north-west and north-east
neighbours of vesicle $x$ are excited. Two wave-fragments enter
vesicle $x$: one fragment travels south-east, another south-west.  The
wave-fragments merge and form a new wave-fragment which travels
south. This fragment collides with the part of vesicle $x$'s wall
lying between south-west and south-east pores. The fragment becomes
extinguished in result of the collision.

If excited neighbours of vesicle $x$ are separated by another
neighbour of $x$ (e.g. vesicles $x_0$ and $x_2$ in
Fig.~\ref{ballsexample}a) then vector of the newly formed wave-fragment
(result of merging two input wave-fragments) points exactly to one
exit pore (e.g. pore connecting vesicle $x$ and vesicle $x_4$ in
Fig.~\ref{ballsexample}a). In example
(Fig.~\ref{typesofwavecollisions}c) north-west and east neighbours of
vesicle $x$ are excited. Two wave-fragments enter vesicle $x$: one
travels south-east another travels west. They merge and form a
wave-fragment travelling south-west. This fragment leaves vesicle $x$
for its southwest neighbour (vesicle $x_4$).

If wave-fragments travel towards each other
(Fig.~\ref{typesofwavecollisions}d) they collide and annihilate in
result of the collision. Thus no excitation leaves vesicle
$x$. Scenarios of collision between three wave-fragments are shown in
\mbox{(Fig.~\ref{typesofwavecollisions}e--g)}.

In scenarios illustrated in (Fig.~\ref{typesofwavecollisions}h--k) at
least two wave-fragments undergo head-on collision and all input
wave-fragment annihilate. The situation
(Fig.~\ref{typesofwavecollisions}h) is similar to
(Fig.~\ref{typesofwavecollisions}c) with the only difference that
three not two waves collide, however the resulting wave-fragment is
the same: it travels south-west and excites south-west neighbour
(vesicle $x_4$) of vesicle $x$.

Configuration of excited neighbours shown in
Fig.~\ref{typesofwavecollisions}f results in annihilation of all three
incoming wave-fragments. Three wave-fragments entering vesicle $x$
travel south-east, south-west and north-west. Wave-fragment travelling
south-east collides head-on with wave-fragment travelling
north-west. Both wave-fragments annihilate as a result of the
collision. At the same wave-fragment travelling south-west collides in
both south-east and south-west fragments.  The south-west travelling
fragment also annihilates (Fig.~\ref{typesofwavecollisions}f).

Situation shown in (Fig.~\ref{typesofwavecollisions}g) does correspond
to direct head-on collision between wave-fragments. Three incoming
wave-fragments travel south-east, west and north-east. A sum of
velocity vectors of these wave-fragments is nil, therefore all
fragments annihilate and no wave-fragments leave vesicle $x$.

In the remaining scenarios (Fig.~\ref{typesofwavecollisions}h--k) at
least two incoming wave-fragments experience head-on collisions and
all wave-fragments annihilate.

\section{Cellular automaton representation}
\label{CAmodel}

We simulate interactions of wave-fragments in BZ-vesicles using
two-dimensional cellular automaton. The cellular automaton is a
hexagonal array $\mathbf L$ of cells. A cell is a finite state machine
which updates its states in discrete time depending on states of its
six closest neighbours. All cells update their states simultaneously.

A cell takes eight states from set ${\mathbf Q}=\{ \circ, -, \searrow,
\swarrow, \leftarrow, \nwarrow, \nearrow, \rightarrow \}$.  They
are~:- resting state ($\circ$), when the cell is ready to be excited,
refractory state ($-$), when the cell does not react to states of its
neighbours, and six excited states representing wave-fragments. The
states representing wave-fragments are coded as $\searrow$ (south-east
travelling wave), $\swarrow$ (south-west travelling wave),
$\leftarrow$ (westward travelling wave), $\nwarrow$ (north-west
travelling wave), $\nearrow$ (north-east travelling wave) and
$\rightarrow$ (eastward travelling wave).  The excited state of a cell
indicates what type of excitation wave-fragment is leaving the cell.

\begin{figure}[!tbp]
\centering
\subfigure[$t=6$]{\includegraphics[width=0.45\textwidth]{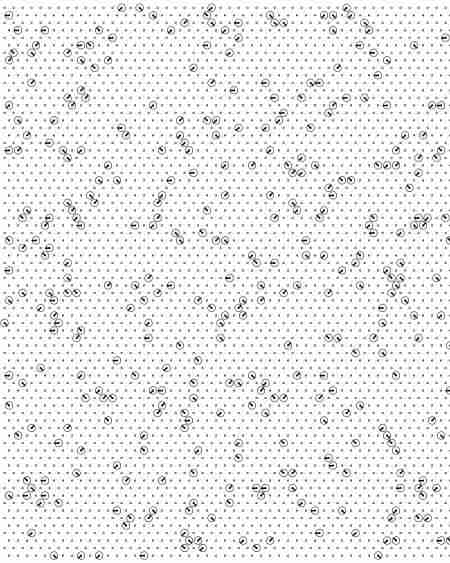}}
\subfigure[$t=7$]{\includegraphics[width=0.45\textwidth]{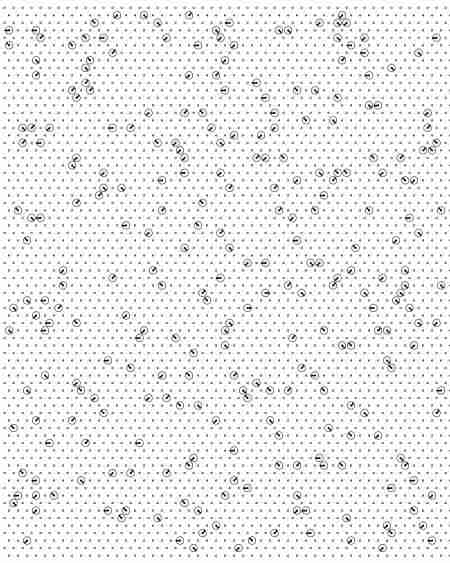}}
\subfigure[$t=8$]{\includegraphics[width=0.45\textwidth]{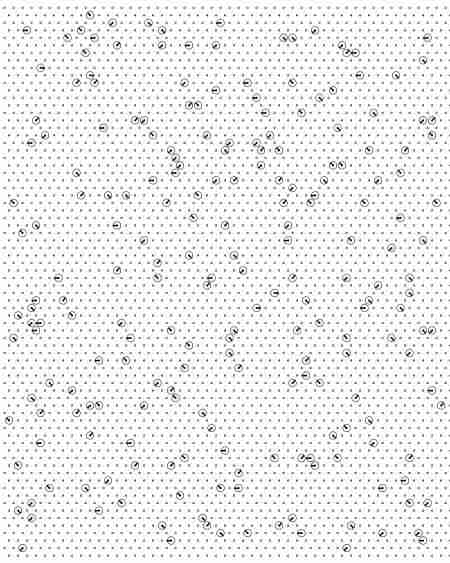}}
\subfigure[$t=9$]{\includegraphics[width=0.45\textwidth]{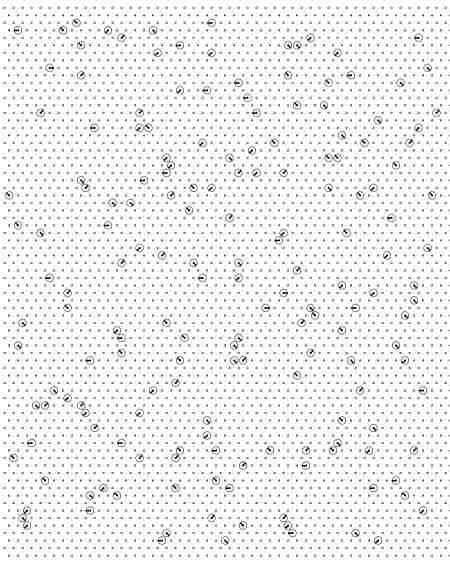}}
\caption{Configurations of the cellular automaton from initially random configuration: a cell gets on of 
six wave-fragment-states with probability 0.1 and gets a resting state otherwise. Boundary are absorbing. 
Resting cell-states are shown by small circles, refractory states are not shown. Cell-states representing 
wave-fragments are shown by large circles with vectors pointed south-east, south-west, west, north-west, north-east,
east.  
}
\label{randomstart}
\end{figure}

Let $s: {\mathbf L} \times {\mathbf Q} \rightarrow \{ 0, 1 \}$ defined as follows. For every $x$ with neighbourhood
$u(x)=\{ x_0, x_1, \ldots, x_5\}$ (Fig.~\ref{ballsexample}a): 
$s(x_0)^t=1$ if $x_0^t=\searrow$ and $s(x_0)^t=0$ otherwise;
$s(x_1)^t=1$ if $x_1^t=\swarrow$ and $s(x_1)^t=0$ otherwise;
$\ldots$;
$s(x_5)^t=1$ if $x_5^t=\rightarrow$ and $s(x_5)^t=0$ otherwise. Cell-state transition rule is defined as follows: 
\begin{equation}
x^{t+1} = 
\begin{cases}
\searrow, & \text{ if } x^t=\circ \text{ and } s(u(x)^t) \in \{ (100000), (010001), (110001) \} \\
\swarrow, & \text{ if } x^t=\circ \text{ and } s(u(x)^t) \in \{ (010000), (101000), (111000) \}\\
\leftarrow, &  \text{ if } x^t=\circ \text{ and } s(u(x)^t) \in \{ (001000), (010100), (011100)\}\\
\nwarrow, &  \text{ if } x^t=\circ \text{ and } s(u(x)^t) \in \{ (000100), (001010), (001110) \}\\
\nearrow, &  \text{ if } x^t=\circ \text{ and }  s(u(x)^t) \in \{ (000010), (000101), (000111) \}\\
\rightarrow, &  \text{ if } x^t=\circ \text{ and } s(u(x)^t) \in \{ (000001), (100010), (100011) \}\\
-, & \text{ if } x^t \in \{ \searrow, \swarrow, \leftarrow, \nwarrow, \nearrow, \rightarrow \} \\
\circ, & \text{ otherwise }.
\end{cases}
\end{equation}

Example of the cellular automaton evolution from initial random
configuration is shown in Fig.~\ref{randomstart}.  In case of
absorbing boundary conditions any random initial configurations leads
to `empty' global configuration where all cells are in resting
states. This is because wave-fragment either annihilate each other (in
majority of local configurations) or produce less new wave-fragments
than those involved in a collision.

\begin{figure}[!tbp]
\centering
\includegraphics[width=0.7\textwidth]{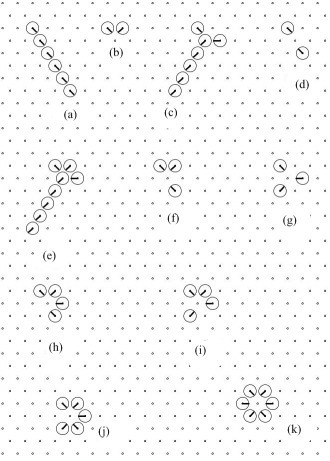}
\caption{Traces of seven steps of cellular automaton development with
  eleven initial sub-configurations corresponding to types of wave
  collisions in Fig.~\ref{typesofwavecollisions}. 
  Cell-states representing 
wave-fragments are shown by large circles with vectors pointed south-east, south-west, west, north-west, north-east,
east.  
  To show the trace of
  the cellular automaton development we do not execute transition to
  refractory state but leave a cell in its wave-fragment state
  instead. Such approach is appropriate when signals trajectories do
  not cross each other.}
\label{caanalogs}
\end{figure}

Examples of collision scenarios of Fig.~\ref{typesofwavecollisions}
represented in cellular automaton configurations are shown in
Fig.~\ref{caanalogs}. The cellular automaton can be used as a `rapid
prototyping' tool to design logical circuits in BZ-vesicle arrays.

\section{Boolean gates}
\label{gatessection}

We adopt principles of collision-based computing~\cite{adamatzky_cbc}
to design logical gates in BZ-vesicle arrays.  We assume travelling
wave-fragments represent values of Boolean variables: presence of a
wave-fragment in a specified site corresponds to logical {\sc True},
absence to logical {\sc False}. When two wave-fragments collide they
annihilate or merge into a new wave-fragment. Newly formed
wave-fragments represent a logical conjunction.  A wave-fragment
travelling along its original (undisturbed) trajectory gives us a
conjunction of a logical variable represented by this wave-fragment
with negation of logical variables represented by another
wave-fragment.

\begin{figure}[!tbp]
\centering
\subfigure[]{\includegraphics[scale=0.6]{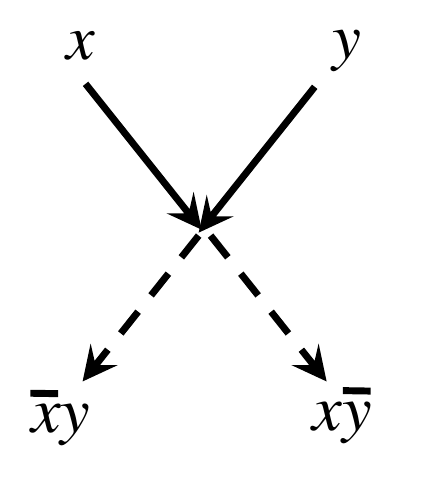}}
\subfigure[]{\includegraphics[scale=0.6]{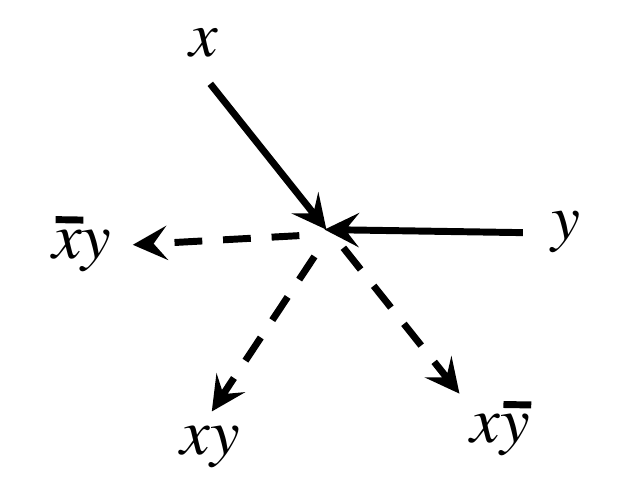}}
\subfigure[]{\includegraphics[scale=0.6]{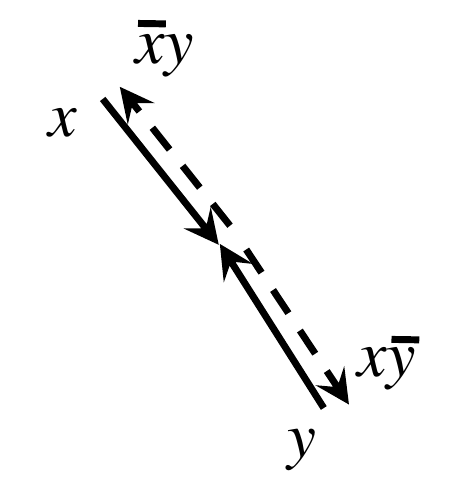}}
\subfigure[]{\includegraphics[scale=0.6]{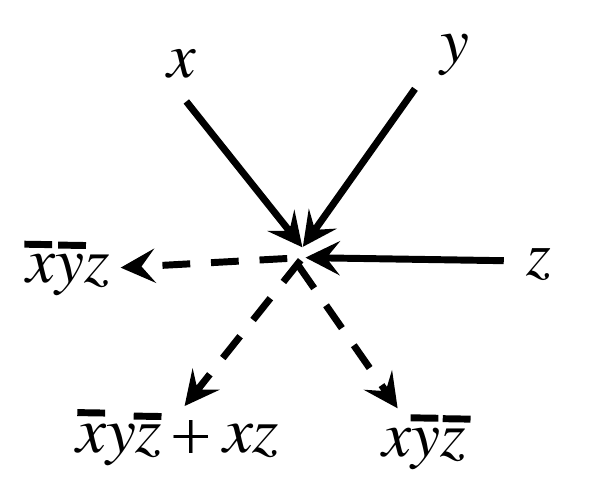}}
\subfigure[]{\includegraphics[scale=0.6]{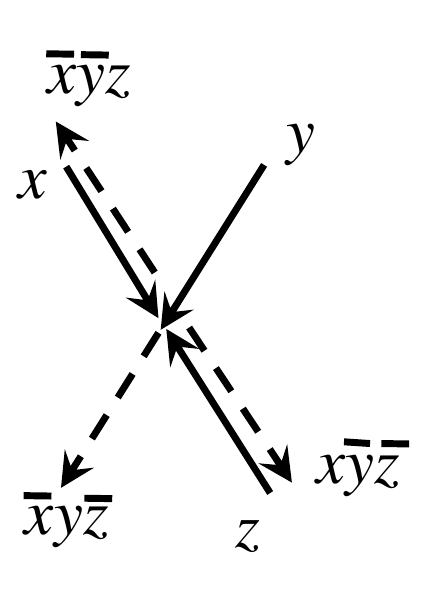}}
\subfigure[]{\includegraphics[scale=0.6]{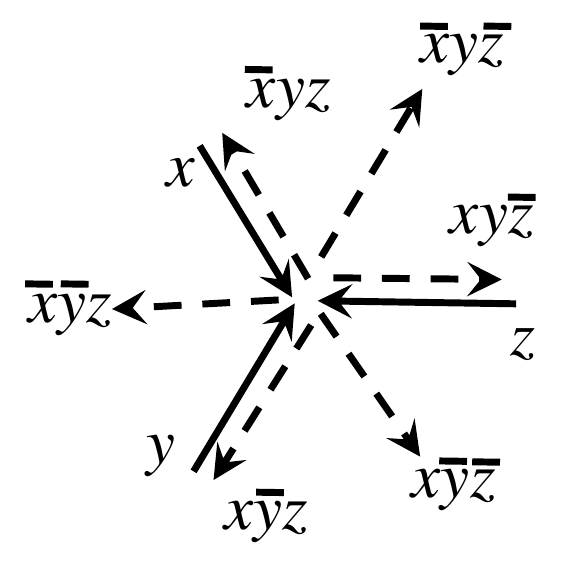}}
\caption{Two and three input Boolean gates implemented in BZ-vesicles
  and the cellular automaton model. Solid lines indicate inputs, dashed lines -- outputs.}
\label{gates}
\end{figure}

With regards to usages of space by incoming and outgoing
wave-fragments collision-based gates can be classified on gates where
input and output trajectories do not overlap (Fig.~\ref{gates}abd) and
gates where we may record output signals from inputs
(Fig.~\ref{gates}abdf).

\subsection{Gates with separate input and outputs}

Gate $\langle x, y \rangle \rightarrow \langle \overline{x}y, x
\overline{y} \rangle$ (Fig.~\ref{gates}a) is implemented in the
situation when wave-fragments incoming to a vesicle originate from the
vesicles which are closest neighbours to each other
(Fig.~\ref{typesofwavecollisions}b and
Fig.~\ref{caanalogs}b). South-east travelling wave-fragment represents
$x$ and south-west travelling wave-fragment represents $y$. If just
one wave-fragment, say $x$, is present then it continues its course
undisturbed, thus its outgoing trajectory symbolises
$x\overline{y}$. When two wave-fragments are present ($x=${\sc true}
and $y=${\sc true}) they collide, form a new wave-fragment 
which annihilate by colliding in a vesicle wall between pores.

In collision scenario (Fig.~\ref{typesofwavecollisions}c and
Fig.~\ref{caanalogs}c) wave-fragments $x$ and $y$ do not annihilate
but form a new wave-fragment which travels along its own
trajectory. This new wave-fragment represents conjunction $xy$. Thus
we have a two-input three-output logical gate $\langle x, y \rangle
\rightarrow \langle x\overline{y}, xy, \overline{x}y \rangle$
(Fig.~\ref{gates}b).

\begin{figure}[!tbp]
\centering
\subfigure[]{\includegraphics[width=0.6\textwidth]{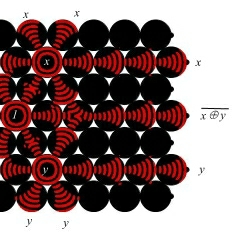}}
\subfigure[]{\includegraphics[width=0.5\textwidth]{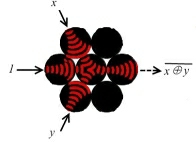}}
\caption{Example of gate $\overline{x \oplus y}$. The figure shows
  time lapse images of propagating excitation wave-fragments:
  (a)~experimental setup with omnidirectional excitation of three
  sites of BZ-vesicle array, (b)~selected local configuration of the
  gate. Constant {\sc True} is shown by '1'.}
\label{gateXoplusY}
\end{figure}

Arrangement of colliding wave-fragments shown in
Fig.~\ref{typesofwavecollisions}e and Fig.~\ref{caanalogs}e implements
three-input three-output gate $\langle x, y, z \rangle \rightarrow
\langle \overline{x}\overline{y}z, \overline{x}y\overline{z}+xz,
x\overline{y}\overline{z} \rangle$ (Fig.~\ref{gates}d). By assigning
constant {\sc True} to one of the inputs we can transform the gate to
negation of exclusive disjunction. An implementation of the gate
$\langle x, y \rangle \rightarrow \langle \overline{x \oplus y}
\rangle$ with collision of wave-fragments is shown in
Fig.~\ref{gateXoplusY}. This example also demonstrates that by
exciting input vesicles with omnidirectional waves we multiply
original signals, and copies of the signals can be employed in further
gates.

\subsection{Gates with spatially overlapping inputs and outputs}

Gate shown in Fig.~\ref{gates}c is the simplest amongst gates with
spatially overlapping trajectories of input and output
wave-fragments. If $y$={\sc False} then no wave-fragment is present in
input $y$. Therefore wave-fragment representing $x=${\sc True} travels
along its original trajectory undisturbed and eventually appears on
the output $x\overline{y}$.  The three-input, three-output gate in
Fig.~\ref{gates}e produces its result $x\overline{y}\overline{z}$,
$\overline{x}\overline{y}z$, $\overline{x}y\overline{z}$ on the output
trajectories co-aligned with input trajectories for signals $x$, $y$
and $z$.

Gate Fig.~\ref{gates}f has three inputs and six outputs. Three outputs
--- $\overline{x}yz$, $xy\overline{z}$, $x\overline{y}z$ --- are
represented by wave-fragments travelling outward the collision site
but along the input trajectories of signals $x$, $z$ and $y$.  Three
other outputs have their own trajectories ---
$x\overline{y}\overline{z}$, $\overline{x}\overline{y}z$, and
$\overline{x}y\overline{z}$.

While it may be convenient to employ gates with spatially overlapping
inputs and outputs, e.g. when space is an issue, a proper functioning
of the gates will require more advanced temporal coordination of
signals.

\section{Towards binary adder}
\label{adder}

We employ some of the gates with spatially separated input and output
trajectories to design components of one-bit binary adder. The example
provided is for illustrative purposes: we did not optimize structure
of logical circuits presented, neither we were concerned with number
of instances of input variables used.

\begin{figure}[!tbp]
\centering
\subfigure[$x=0, y=0, z=0$]{\includegraphics[width=0.27\textwidth]{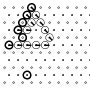}}
\subfigure[$x=0, y=0, z=1$]{\includegraphics[width=0.27\textwidth]{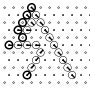}}
\subfigure[$x=0, y=1, z=0$]{\includegraphics[width=0.27\textwidth]{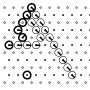}}
\subfigure[$x=0, y=1, z=1$]{\includegraphics[width=0.27\textwidth]{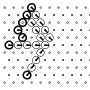}}
\subfigure[$x=1, y=0, z=0$]{\includegraphics[width=0.27\textwidth]{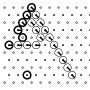}}
\subfigure[$x=1, y=0, z=1$]{\includegraphics[width=0.27\textwidth]{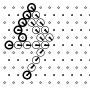}}
\subfigure[$x=1, y=1, z=0$]{\includegraphics[width=0.27\textwidth]{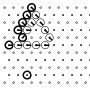}}
\subfigure[$x=1, y=1, z=1$]{\includegraphics[width=0.27\textwidth]{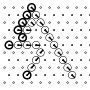}}
%\subfigure[Circuit scheme]{\includegraphics[scale=0.4]{figs/x_XOR_y_XOR_cin_gate.pdf}}
\caption{Calculation of sum $S=(x \oplus y) \oplus z$, $z=c_{in}$, in a cellular automaton model 
of BZ-vesicle hexagonal array: (a)--(h)~traces of cell-states, representing wave-fragments, for 
all possible combinations of input values (Fig.~\ref{adder_scheme}). Input cells are shown by circles with thick lines.}
\label{sum}
\end{figure}

\begin{figure}[!tbp]
\centering
\subfigure[$x=0, y=0, z=0$]{\includegraphics[width=0.27\textwidth]{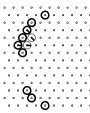}}
\subfigure[$x=0, y=0, z=1$]{\includegraphics[width=0.27\textwidth]{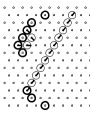}}
\subfigure[$x=0, y=1, z=0$]{\includegraphics[width=0.27\textwidth]{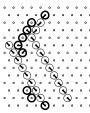}}
\subfigure[$x=0, y=1, z=1$]{\includegraphics[width=0.27\textwidth]{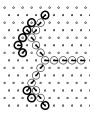}}
\subfigure[$x=1, y=0, z=0$]{\includegraphics[width=0.27\textwidth]{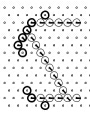}}
\subfigure[$x=1, y=0, z=1$]{\includegraphics[width=0.27\textwidth]{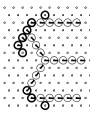}}
\subfigure[$x=1, y=1, z=0$]{\includegraphics[width=0.27\textwidth]{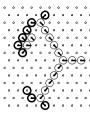}}
\subfigure[$x=1, y=1, z=1$]{\includegraphics[width=0.27\textwidth]{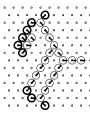}}
%\subfigure[Circuit scheme]{\includegraphics[scale=0.4]{figs/CarryOut_gate.pdf}}
\caption{Calculation of carry out value $C_{out}=xy + z(x \oplus y)$, $z=c_{in}$ in 
cellular automaton model of BZ-vesicle hexagonal array: (a)--(h)~traces of cell-states, representing wave-fragments, for 
all possible combinations of input values (Fig.~\ref{adder_scheme}). Input cells are shown by circles with thick lines.
Eastward travelling wave-fragments in 3rd and 13th rows in (ef) are unused byproducts of the circuit.}
\label{carryout}
\end{figure}

\begin{figure}[!tbp]
\centering
\subfigure[Sum ($S$)]{\includegraphics[scale=0.4]{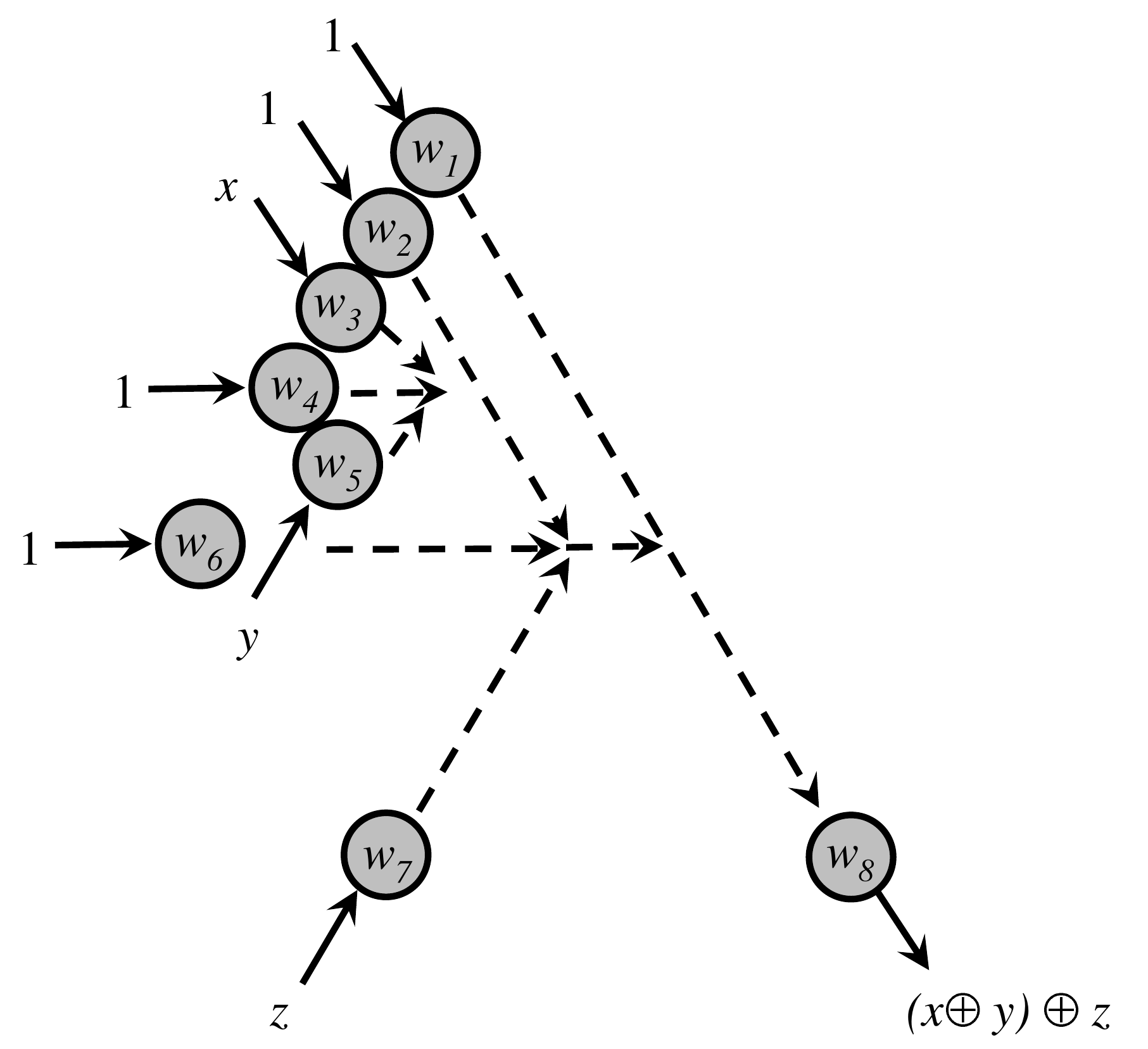}}
\subfigure[Carry out ($C_{out}$)]{\includegraphics[scale=0.4]{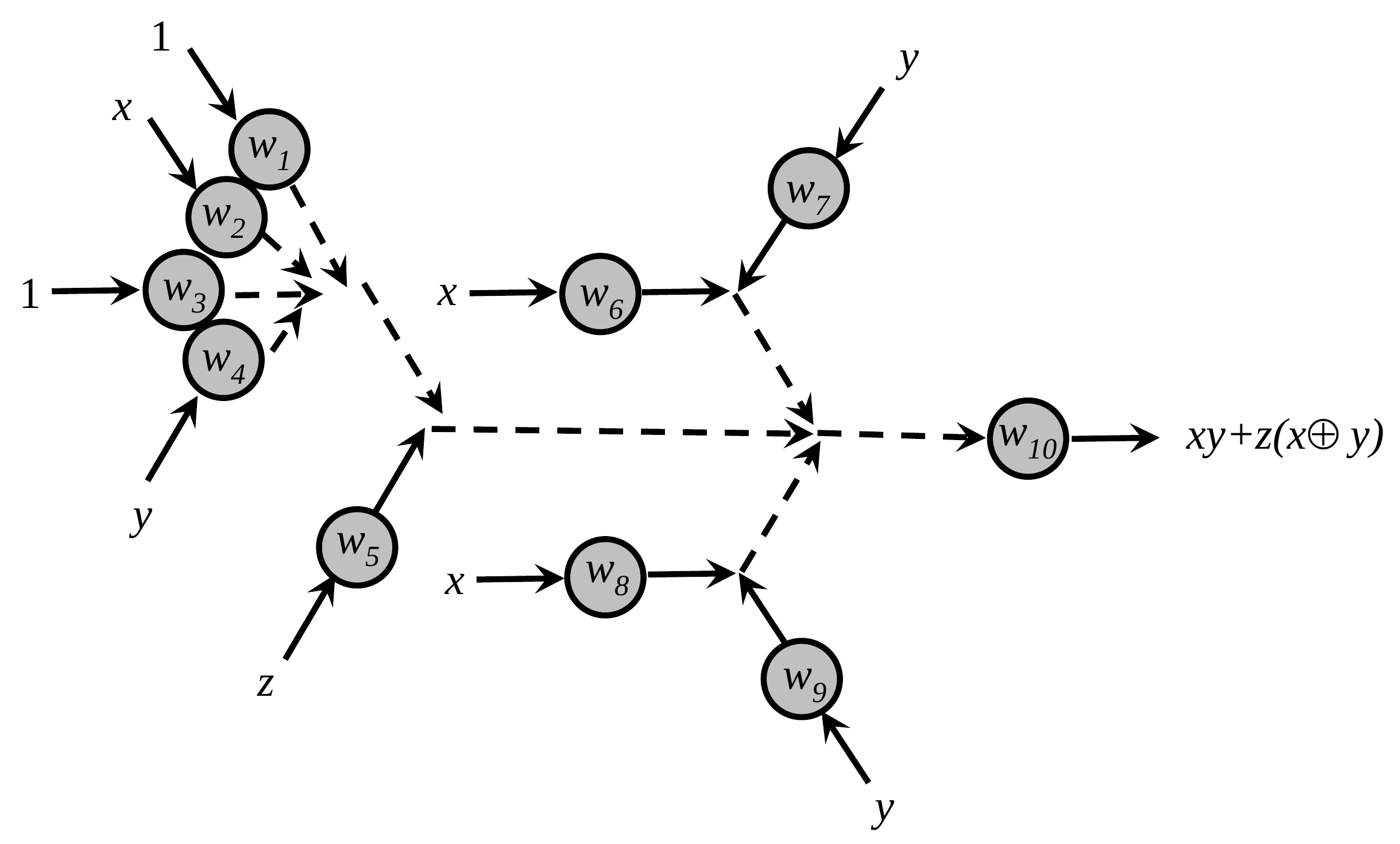}}
\caption{Scheme of the circuit. (a)~Sum (Fig.~\ref{sum}) \& (b)~Carry
  out (Fig.~\ref{carryout}).}
\label{adder_scheme}
\end{figure}

A binary adder based on BZ-vesicles uses collisions between
propagating wave-fragments to perform addition of three one-bit binary
numbers $x$, $y$ and $z$. The signal $z$ depict $C_{in}$.  There are
many versions of particular implementations of full one-bit adder.  We
adopted the most common one, where the adder outputs sum of signals
$S= (x \oplus y) \oplus z$ (Fig.~\ref{sum} \& Fig.~\ref{adder_scheme}a)
and carry out value $C_{out}= xy + z(x \oplus y)$
(Fig.~\ref{carryout} \& Fig~\ref{adder_scheme}b).

Positions of inputs in Figs.~\ref{sum} and \ref{carryout} are shown by
thick solid circles. If an input equals {\sc False} the circle
contains only central dot, if the input equals {\sc True} the circle
contains a vector indicate the sites state. Result of sum circuit is
represented by wave-fragment travelling south-east while output of
carry out circuit is represented by wave-fragment travelling east. The
circuit calculating $S$ is packed in array of $6 \times 10$
BZ-vesicles, and the circuit calculating $C_{out}$ occupies as
sub-array of $7 \times 13$ BZ-vesicles. Let us discuss functioning of 
circuit $S$ (Fig.~\ref{sum}) in details.  

\subsection{Sum}

If all three inputs are {\sc False} no waves are initiated in sites $w_3$ and $w_5$ and $w_7$
(Fig.~\ref{sum}a and \ref{adder_scheme}a). The wave-fragments representing constant
{\sc True} are placed in sites $w_1$  and $w_2$ (south-east travelling wave-fragments), 
$w_4$ and $w_6$ (wave-fragments travelling east). Wave-fragment $w_2$ collides with wave-fragment $w_4$, both 
wave-fragments annihilate (step three of cellular automaton simulation, Fig.~\ref{sum}a). 
Wave-fragments $w_1$ and $w_6$ travel a bit further but collide one with another and annihilate at
fifth step of automaton simulation. All outputs are {\sc False} therefore. 

If carry in value $z$ is {\sc True} and other input values are {\sc False} (Fig.~\ref{sum}a and \ref{adder_scheme}b)  
then north-east traveling wave $w_7$ annihilates east travelling wave $w_6$. Therefore wave $w_1$ continues 
travelling south-east, thus representing {\sc True} value of $(x \oplus y) \oplus z$.  

In situation $(x, y, z) = (${\sc False, True, False}$)$ wave $w_5$ collides with wave $w_4$ (Fig.~\ref{sum}c and \ref{adder_scheme}a), both wave-fragments annihilate. Therefore wave-fragment $w_2$ continues undisturbed its travel to 
south-east where it collides with wave-fragment $w_6$. With both waves representing east travelling constant
{\sc True} cancelled  wave-fragment $w_1$ reaches an output. Similarly for input 
$(x, y, z) = (${\sc True, False, False}$)$ wave-fragments $w_3$ and $w_4$, and $w_2$ and $w_6$ 
annihilate in collision with each other (Fig.~\ref{sum}e and \ref{adder_scheme}a). Therefore wave-fragment $w_1$ travels undisturbed. Other combinations of inputs can  be considered in the similar manner.

\subsection{Carry out}

Computation starts in a group of sites marked $w_1, \ldots w_4$ 
in (Fig.~\ref{adder_scheme}b): site $w_2$ represents $x$, site $w_4$ represents $y$, and sites $w_1$  and $w_3$ represent 
constant {\sc True}. The sub-circuits' output is $x \oplus y$. Wave-fragments $w_1$, travelling south-east, and $w_3$, travelling east, are always present in the system. If only one of inputs $x$ or $y$ has {\sc True} value, e.g. south-east travelling wave-front $w_2$ in Fig.~\ref{carryout}ef), then wave-fragment representing this input collides with wave-fragment $w_3$ (constant {\sc True}) and both wave-fragments annihilate. If both inputs $x$ and $y$ are {\sc True} (Fig.~\ref{carryout}gh) 
then waves $w_2$ and $w_4$ collide with each other and wave $w_3$, merge altogether and produce new wave travelling east. 
This new wave (seen in node $(6,5)$ in (Fig.~\ref{carryout}gh)) collides with wave $w_1$, both wave-fragments annihilate.    
Thus, sub-circuit $(w_1, \ldots, w_4)$ computes $x \oplus y$.

Output wave-fragment of sub-circuit $x \oplus y$ collides with wave-fragment $w_5$, which represents carry in value $z$. These wave-fragments collide at angle $120^o$, therefore they merge and produce new wave-fragment (travelling east) when collide.
See examples in Fig.~\ref{carryout}df. Thus intermediary result $z(x \oplus y)$ is calculated. 

Two small sub-circuits --- $w_6$ and $w_7$, and $w_8$ and $w_9$ -- are arranged symmetrically north and 6,6south of trajectory of wave-fragment, which represents $z(x \oplus y)$. Each of the sub-circuits $(w_6, w_7)$ and $(w_8,w_9)$ produces $xy$. The final
wave-fragment $w_10$ is only produced when either only wave-fragment $z(x \oplus y)$ travels east (Fig.~\ref{carryout}df), or all three wave-fragments ---
wave-fragment $xy$ (output of sub-circuits $(w_6,w_7)$) travelling south-east,
wave-fragment $xy$  (output of sub-circuits $(w_8,w_9)$) travelling north-east, and 
wave-fragment $z(x \oplus y)$ collide (Fig.~\ref{carryout}h).

%\clearpage

\section{Summary}
\label{discussion}

Belousov-Zhabotinsky (BZ) medium in a sub-excitable mode exhibit the
rich spatio-temporal dynamics of mobile self-localisations,
wave-fragments behaving as distant analogs of dissipative solitons. In
an ideal situation a wave-fragment keeps its shape and travels in a
predetermined direction until a collision with another
wave-fragment. Therefore wave-fragments can be used as a quanta of
information, or analogs of billiard
vesicles~\cite{fredkin_toffoli_1982}, in developing collision-based
computing circuits~\cite{adamatzky_cbc}. Our attempts to implement
large scale logical circuits in experimental chemical laboratory
conditions show that wave-fragments are very unstable and do not
conserve their shape for a long time without external
control~\cite{RITABEN,ben_gun}.

To overcome the problem of wave-fragment instability we decided to
achieve a compromise between geometrically-constrained and free space
approaches in designing reaction-diffusion chemical computers. In
computer simulation experiments we designed a lattice of
two-dimensional vesicles (BZ-vesicles) filled with BZ reaction
solution. The BZ-vesicles are in close contact with each other, via
tiny pores, and thus excitation wave can pass form one vesicle to its
closest neighbour. Wave-fragments keep their shape or just slightly
expand inside each BZ-vesicle. When a wave-fragment passes from one
vesicle to another it contracts, due to the restricted size of the
connecting pore.

In computer experiments we demonstrated that each BZ-vesicle, in a hexagonal 
arrangement of vesicles, can act
as a logical gate with at most six inputs and six outputs. All basic
operations of a Boolean logic are implemented via collision between
wave-fragments in a vesicle and thus a lattice of BZ-vesicles is a
computationally universal substrate. To show that BZ-vesicles pose not
only a purely theoretical interest we cascaded the logical gates,
mostly gates with two inputs and one or two outputs, in logical
circuits which implement addition of two one-bit binary
numbers. Functioning of the circuits was verified in
cellular-automaton models of BZ-vesicle array.

Implementation of theoretical designs in chemical laboratory
experiments would be the next step of our studies in fine-grained
compartmentalised excitable chemical systems. Some experimental works
have been done, Kitahata and colleagues provided amazing experimental
evidences of excitation wave-travelling inside a tiny sphere and even
movement of BZ-vesicle partly controlled by wave
propagation~\cite{kitahata_2002, kitahata_2006}. Also, BZ reaction
mixture can be encapsulated via dispersion of the mixture in a
water-in-oil micro-emulsion with surfactant: BZ reagents becomes
enclosed in a mono-layer of anionic surfactant. `Unlimited' energy supply 
from solution surrounding BZ-vesicles enables reactivity of BZ droplets 
to be sustained for extended periods of time~\cite{epstein_2005}.

\section{Acknowledgements}

The work is part of the European project 248992 funded under 7th FWP
(Seventh Framework Programme) FET Proactive 3: Bio-Chemistry-Based
Information Technology CHEM-IT (ICT-2009.8.3). We thank the project
coordinator Peter Dittrich and project partners Jerzy Gorecki and
Klaus-Peter Zauner for their inspirations and useful discussions.


\begin{thebibliography}{99}

\bibitem{adamatzky_2004_collision} Adamatzky A. Collision-based
  computing in Belousov--Zhabotinsky medium. Chaos Solitons Fractals
  21 (2004) 1259--1264.

\bibitem{adamatzky_cbc} Adamatzky~A. (Ed.)  Collision Based
  Computing. Springer, 2002.

\bibitem{adamatzky_delacycostello_asai_2005} A.~Adamatzky, B.~De Lacy
  Costello, and T.~Asai, Reaction Diffusion Computers (Elsevier,
  2005).

\bibitem{andy_ben_BZ_collision} Adamatzky A., and De Lacy Costello B.
  Binary collisions between wave-fragments in a sub-excitable
  Belousov-Zhabotinsky medium. Chaos, Solitons \& Fractals 34 (2007)
  307--315.

\bibitem{adamatzky_physarumgates} Adamatzky~A.  Slime mould logical
  gates: exploring ballistic approach.  arXiv:1005.2301v1 [nlin.PS]
  (2010).  \url{http://arxiv.org/abs/1005.2301}

\bibitem{adamatzky_privman_issue} Adamatzky~A.  Topics in
  reaction-diffusion computer.  J Computational and Theor Nanoscience.
  (2010), in press.

\bibitem{beato_engel} Beato~V., Engel~H. Pulse propagation in a model
  for the photosensitive Belousov-Zhabotinsky reaction with external
  noise. In: Noise in Complex Systems and Stochastic Dynamics, Edited
  by Schimansky-Geier~L., Abbott~D., Neiman~A.,
  Van~den~Broeck~C. Proc. SPIE 5114 (2003) 353--362.

\bibitem{ben_gun} De~Lacy~Costello~B., Toth~R., Stone~C.,
  Adamatzky~A., Bull~L.  Implementation of glider guns in the
  light-sensitive Belousov-Zhabotinsky medium Phys. Rev. E 79 (2009)
  026114 .

\bibitem{epstein_2005} Epstein~I.~R. and Vanag~V.~K.  Complex patterns
  in reactive microemulsions: self-organized nanostructures?  Chaos 15
  (2005) 047510.

\bibitem{field_noyes_1974} Field~R.~J., Noyes~R.~M.  Oscillations in
  chemical systems. IV. Limit cycle behavior in a model of a real
  chemical reaction.  J. Chem. Phys. 1974 (60) 1877--1884.

\bibitem{fredkin_toffoli_1982} Fredkin~E. and Toffoli~T.  Conservative
  logic.  Int J Theor Physics 21 (1982) 219-–253.

\bibitem{gorecka_2003} G\'{o}recka J. N., G\'{o}recki J.  T-shaped
  coincidence detector as a band filter of chemical signal frequency,
  Phys. Rev. E 67 (2003) 067203.

\bibitem{gorecki_2003} G\'{o}recki J., Yoshikawa K. and Igarashi Y.,
  On chemical reactors that can count, J. Phys. Chem. A 107 (2003)
  1664--1669.

\bibitem{gorecki_2005} G\'{o}recki J., G\'{o}recka J. N., Yoshikawa
  K., Igarashi Y., Nagahara H.  Sensing the distance to a source of
  periodic oscillations in a nonlinear chemical medium with the output
  information coded in frequency of excitation pulses. Phys. Rev. E 72
  (2005) 046201.

\bibitem{gorecki_2006} G\'{o}recki J. and G\'{o}recka J. N.,
  Multi-argument logical operations performed with excitable chemical
  medium, J. Chem. Phys. 124 (2006) 084101.

\bibitem{gorecki_2006a} G\'{o}recki J., G\'{o}recka J. N.  Information
  processing with chemical excitations --- from instant machines to an
  artificial chemical brain Int J Unconv Comput 2 (2006) 321--336.

\bibitem{gorecki_2009} G\'{o}recki~J., G\'{o}recka~J.~N., Igarashi~Y.
  Information processing with structured excitable medium, Natural
  Computing 8 (2009) 473--492.

\bibitem{gorecka_2007} G\'{o}recka J. N., G\'{o}recki J., Igarashi Y.
  On the simplest chemical signal diodes constructed with an excitable
  medium, Int J Unconventional Computing 5 (2009) 129--143.

\bibitem{gorecki_private} Gorecki~J. Private communication (2010).

\bibitem{kaminaga_2006} Kaminaga~A., Vanag~V.~K., and Epstein~I.~R.  A
  reaction--diffusion memory device.  Angew. Chem. Int. Ed. 45 (2006)
  3087-–3089.

\bibitem{kitahata_2002} Kitahata~H., Aihara~R., Magome~N., and
  Yoshikawa~K.  Convective and periodic motion driven by a chemical
  wave.  J. Chem. Phys. 116 (2002) 5666--5672.

\bibitem{kitahata_2006} Kitahata~H.  Spontaneous motion of a droplet
  coupled with a chemical reaction.  Prog. Theor. Phys. Suppl. 161
  (2006) 220--223.

\bibitem{motoike_2003} Motoike~I.~N. and Yoshikawa~K.  Information
  operations with multiple pulses on an excitable field.  Chaos,
  Solitons \& Fractals 17 (2003) 455--461.

\bibitem{neuneu} NeuNeu: Artificial Wet Neuronal Networks from
  Compartmentalised Excitable Chemical Media.  (2010)
  \url{http://neu-n.eu/}

\bibitem{sakurai_2002} T. Sakurai, E. Mihaliuk, F. Chirila, and
  K. Showalter, Design and control of wave propagation patterns in
  excitable media, Science 296 (2002) 2009--2012.

\bibitem{sielewiesiuk_2001} Sielewiesiuk J. and G\'{o}recki J.,
  Logical functions of a cross junction of excitable chemical media,
  J. Phys. Chem., A105 (2001) 8189.

\bibitem{RITABEN} Toth~R., Stone~C., Adamatzky~A., de Lacy
  Costello~B., Bull~L.  Experimental validation of binary collisions
  between wave-fragments in the photosensitive Belousov-Zhabotinsky
  reaction.  Chaos, Solitons \& Fractals 41 (2009) 1605--1615.
  
\bibitem{toth_2009}  
  Toth~R., Stone~C., De Lacy Costello~B., Adamatzky~A., Bull~L. 
  Simple collision-based chemical logic gates with adaptive computing. 
  J Nanotech and Molecular Computation 1 (2009) 1-13.

\bibitem{yoshikawa_2009} Yoshikawa~K., Motoike~I.~M., Ichino~T.,
  T. Yamaguchi, Y. Igarashi, J. Gorecki and J. N. Gorecka Basic
  information processing operations with pulses of excitation in a
  reaction-diffusion system.  Int J Unconventional Computing 5 (2009)
  3--37.
 
\bibitem{yoshikawa_2009a} Yoshikawa~K., Nagahara~H., Ichino~T.,
  J. Gorecki, J. N. Gorecka and Y. Igarashi On chemical methods of
  direction and distance sensing.  Int J Unconventional Computing 5
  (2009) 53--65.

\end{thebibliography}
\end{document}